\documentstyle[12pt]{article}

\font\twelve=cmbx10 at 15pt
\font\ten=cmbx10 at 12pt

\begin{document}

\begin{titlepage}

\begin{center}

{\ten Centre de Physique Th\'eorique - CNRS - Luminy, Case 907}

{\ten F-13288 Marseille Cedex 9 - France }

{\ten Unit\'e Propre de Recherche 7061}

\vspace{2 cm}

{\twelve INTRODUCTION TO STRONG HIGGS SECTOR}

\vspace{0.3 cm}

{\bf Pierre CHIAPPETTA}

\vspace{4 cm}

{\bf Abstract}

\end{center}

A brief introduction to strong Higgs sector i.e. the possibility of
breaking electroweak symmetry without an elementary Higgs is given.
Constraints from present LEP data are studied and the discovery
potential of future colliders investigated.

\vspace{4,5 cm}


\bigskip

\noindent April 1994

\noindent CPT-94/P.3026

\bigskip

\noindent anonymous ftp or gopher : cpt.univ-mrs.fr

\end{titlepage}

\section{Introduction}

\par\noindent The problem of symmetry breaking in electroweak
interactions is achieved in the Standard Model (hereafter denoted as
SM) by adding to the $SU(2)_L \otimes U(1)_Y$ gauge theory the Higgs
lagrangian~\cite{1}~:
$$D_{\mu}\Phi {D^{\mu}\Phi}^{\dagger} - V(\Phi) $$
where
$$V(\Phi) = -{\mu}^2 {\mid \Phi \mid}^2 + \lambda {\mid \Phi
\mid}^4,$$
$\Phi$ being a $SU(2)$ doublet.

IF ${\mu}^2$ is negative $V(\Phi)$ has a minimum at~:
$$\langle \Phi \rangle = \sqrt{\left({-\mu^2 \over \lambda}\right)} =
v = 246\ GeV\, .$$

This mechanism gives masses to the weak bosons $W$ and $Z$ and also
to quarks and leptons through Yukawa couplings. Since the Higgs
doublet has four degrees of freedom and three are needed to give
masses to weak bosons one is left with a scalar boson not yet
discovered which is heavier than $63.5\ GeV$ as indicated by LEP
measurements~\cite{2}.

Even if we can prove that the SM is not trivial i.e. that the Higgs
self coupling constant $\lambda$ has not to be zero, which is still an
open problem~\cite{3}, this mechanism of symmetry breaking is
nevertheless valid up to some scale $\Lambda$, we will specify. In
fact the Higgs part of the lagrangian is not asymptotically free and
the Landau pole, which corresponds to the scale where the coupling
$\lambda$ is infinite, is very low (in the $TeV$ range). The situation
is more complicated for the SM since it involves also  fermions. It
has been shown that if one assumes that the SM is valid up to a grand
unification scale the Higgs has to be
light~\cite{4}. Moreover in order to remain perturbative, the Higgs
cannot be heavier than $630\ GeV$ indicating that
$\Lambda$ is around the $TeV$ scale. Moreover if a light Higgs is
discovered, some new physics is mandatory to control radiative
corrections.

In what follows we are interested in a scenario where no elementary
Higgs boson exists. The Symmetry breaking mechanism is viewed as a
manifestation of some underlying unknown strong interaction at
the $TeV$ scale. In what follows we will not consider a specific
model like technicolor or extended technicolor~\cite{5} but describe
a general framework involving symmetries~\cite{6}. More precisely the
techniques we will use are based on the breaking of a  global
symmetry and similar to the description of low energy strong
interactions involving pions in terms of effective theories ($\sigma$
models) valid up to the
$GeV$ scale.

In analogy with strong interactions the pions correspond to
longitudinal gauge bosons and we will build vector resonances (like
the $\rho$).

We will first show that the lagrangian of the SM can be viewed as the
lagrangian of a linear $\sigma$ model when gauge interactions are put
to zero. The Higgs doublet can be  rewritten in a matricial notation~:
$$M = {\sqrt 2} \left(i {\tau}_2 \Phi^{\star}, \Phi\right)$$
$$L = {1 \over 4} Tr \left(\partial_{\mu} M
\partial^{\mu} M^{\dagger}\right)-{\lambda \over
4}\left({1\over 2} Tr (M M^{\dagger}) + {\mu^2
\over \lambda}\right)^2$$

To exhibit the correspondance with the linear $\sigma$ model we
express the SU(2) matrix as~:
$$M = \sigma + i \pi. \tau $$

The lagrangian obtained has a global  invariance under $SU(2)_L
\otimes SU(2)_R $
$$ M \longrightarrow {\exp\ \left(i {\epsilon}_L {\tau
\over 2}\right)} M {\exp\ \left(-i{\epsilon}_R {\tau
\over 2}\right)}$$
which is spontaneously broken to $SU(2)_V$ if ${\mu}^2$ is negative.
We get~:
$$ L = \left({\partial_{\mu} \sigma} {\partial^{\mu}
{\sigma}}\right)  + \left({\partial_{\mu} \pi} {\partial^{\mu}
{\pi}}\right) + {\lambda\over 4} {\left(\pi^2 + \sigma^2 -
v^2\right)}^2 $$

Let us now take the limit $\lambda \rightarrow \infty $. We must
have~:
${(\pi^2 + \sigma^2 - v^2)}^2 = 0$, which allows to eliminate the
$\sigma$ i.e. the Higgs field.

In this limit the Higgs part of the SM corresponds
to a non linear $\sigma$ model~:
$$ L = {v^2 \over 4} Tr \left({\partial_{\mu} U}
{\partial^{\mu} U^{\dagger}}\right)$$
with the condition $ U {U^{\dagger}} =1 $.

\noindent $U = \displaystyle {M \over v}$ belongs to the coset space
$SU(2)_L\otimes SU(2)_R / SU(2)_V$ and the degrees of freedom are the
Goldstone bosons.

This lagrangian is the basic ingredient for the evaluation of
scattering among longitudinal gauge bosons when the Higgs boson is
very massive. These low energy theorems~\cite{7} are similar to those
describing interactions among pions. They are valid in the energy
range~:
$$ M_W \ll \sqrt{S} \ll {4 \sqrt{\pi} v}\, , $$
the upper limit being given by unitarity.

The upper limit may be lower if for example a vector resonance
exists~: this is indeed the case in strong interactions due to the
existence of the $\rho$. The experimental signal will  consist in an
excess of production of pairs of longitudinal gauge bosons, difficult
to be seen experimentally~\cite{8}.

\section{Vector resonances}

The rest of the lecture is devoted to the description of vector
resonances. The basic idea is to build the equivalent of the  $\rho$
resonance using the concept of hidden symmetry~\cite{9}. It is based
on the fact that non linear $\sigma$ models acting on a coset space
${G \over H}$ can be formulated as gauge theories of a
group $ \bf H $ which is the local version of the residual group H
after symmetry breaking. We are therefore looking for a lagrangian
invariant under~:
$$\left[ {SU(2)_L \otimes SU(2)_R} \right]_{GLOBAL} \otimes
{SU(2)_V}_{LOCAL}\, .$$

If $g$ is an element of the global group G and h an element of $\bf
H$ we have the coset decomposition~:
$$g = \xi h\, .$$

We now introduce the Maurer Cartan differential form~:
$$ {\omega}_{\mu} = {{{\xi}^{\dagger}} \partial_{\mu} {\xi}}$$
that we decompose in a component belonging to $\bf H$ called $
{\omega}_{\mu}^{\parallel}$ and an orthogonal one
${\omega}_{\mu}^{\perp}$ belonging to the coset space.

The invariant quantities we can construct, which will lead to the
most general lagrangian, are~:
$$L^{(1)} = Tr \left( {\omega}_{\mu}^{\perp}
{\omega}^{\mu\perp}\right)$$
and
$$L^{(2)} = \alpha Tr\left({\omega}_{\mu}^{\parallel}
{\omega}^{\mu\parallel}\right),$$
$\alpha$ being an arbitrary parameter.

We recover the SM by the gauge choice $g(x) = {\xi}(x)$, which is
equivalent to a gauge fixing term in Yang Mills theories.

Under local gauge transformations $ {\omega}_{\mu}^{\parallel}$
becomes~:
$$ {\omega}_{\mu}^{\parallel} \longrightarrow { {h^{\dagger}}
{\omega}_{\mu}^{\parallel} h} + {h^{\dagger} {\partial_{\mu}} h}$$
i.e. acts as a triplet of gauge bosons $V_{\mu}$~:
$$V_{\mu} \longrightarrow {2 \over g^{\prime \prime}}
h^{\dagger}\partial_{\mu} h + h^{\dagger} V_{\mu} h\, ,$$
$g^{\prime \prime}$ being the $SU(2)_V$ coupling constant.

If no kinetic term for $V_{\mu}$ is present it is an auxiliary
field and the lagrangian contains only the $L^{(1)}$ piece.

The main assumption is that $V_{\mu}$ has to be a dynamical field.
We will therefore add a kinetic term to the lagrangian~:
$$ L^{kin} = - \left( {1 \over g^{\prime \prime}} \right)^2 F_{\mu
\nu} (V) F^{\mu \nu} (V)\, .$$

Then we have to perform an $SU(2)_L \otimes U(1)_Y \otimes SU(2)_V$
gauging and to eliminate goldstone bosons  in order to give masses to
$W,Z$ and $V$ bosons by going into the unitary gauge. One obtains
finally a  non renormalisable gauge theory  with gauge group~:
$SU(2)_L
\otimes U(1)_Y \otimes SU(2)_V$ which breaks into
$U(1)_{EM}$. The lagrangian we have built is called minimal
BESS~\cite{6},~\cite{10} and predicts the existence of a triplet of
gauge bosons. It contains three parameters~:

\begin{itemize}

\item[---] the mass $M_V$ of the V triplet~: $M_V^2 = \alpha
\displaystyle{v^2 \over 4} g^{\prime\prime}$.

\item[---] the $SU(2)_V$ coupling constant $g^{\prime\prime}$ (we
recover the SM in the limit ${g^{\prime\prime}}\rightarrow \infty$)

\item[---] a direct coupling of $V$ bosons to fermions hereafter
called
$b$. The coupling arises naturally from mixing.

\end{itemize}

Since $W^{\pm}$ mixes with $V^{\pm}$ and $Z^0$ with $V^0$, present
colliders  and especially LEP are sensitive to the strong Higgs
sector. The existence of $V$ bosons affects not only the masses of
$Z^0$ and $W$ bosons but also their couplings to fermions and the
trilinear ones. The order of magnitude of the mixing angles and of the
mixing parameter $b$ is~:
$${g \over g^{\prime \prime}}\cdotp $$

LEP 200 wil not improve significantly present LEP limits which are
displayed in ref.~\cite{11}. We will to wait for linear $e^+e^-$ colliders
and to look for the process $e^+e^-\rightarrow W^+W^-$ well
sensitive to $V^0$~\cite{11}, provided $W$ polarisation
will be measurable.

LHC is well suited for $V^{\pm}$ identification through the reaction
$pp \rightarrow WZ$ by looking at the invariant $WZ$ mass
distribution or for a jacobian peak in the transverse momentum of the
$Z$~\cite{12}. It allows to discover charged
vector resonances up to $2\ TeV$.

Extended technicolor is a particular case of the BESS model. It is
assumed that the underlying strong interaction responsible  for
symmetry breaking is a scaled QCD scenario of group $SU(N)_{TC}$
involving  doublets of technifermions whose condensates break the
symmetry like ordinary quarks condensates break the  chiral symmetry
in strong interactions. In extended technicolor the parameter
$\alpha$  is fixed to $2$ and $N_{TC}$ is related to $g^{\prime
\prime}$.

The direct coupling to fermions is also fixed~:
$$b = -2 \left({v \over \Lambda_{ETC}}\right)^2$$
where $\Lambda_{ETC}$ is the unknown scale for extended technicolor.

LEP has already excluded~:
$$N_{TC} N_{DOUBLETS} \leq 12$$

The BESS model can be extended~\cite{13} if we start from a larger
global symmetry like ${SU(8)_L} \otimes {SU(8)_R}$  broken into
${SU(8)_{L+R}}$. This leads to a very rich spectrum  of vector
reso\-nances, axial vector ones and also pseudogoldstone bosons which
are either singlets or triplets of $SU(2)_L$ and either
singlets, triplets or octets of $SU(3)_C$.

Since they give a negative contribution to one of the self
energies  of the SM, i.e. ${\epsilon}_3$, the extended BESS model is
still alive for a larger parameter space domain. Moreover the bounds
on the top mass, which can be derived from ${\epsilon}_1$, are
weakened. This self energy is also sensitive to the splitting in
pseudogoldstone multiplets. Compared to minimal BESS the width of
vector resonances increases and the identification of $V^{\pm}$
throught $WZ$ channel is less promising. The production of
pseudogoldstone bosons at hadronic colliders suffers from a huge
hadronic background. Linear $e^+ e^-$ colliders are more promising if
resonant production from $V^0$ is possible.

\section{Conclusion}

To conclude present LEP data constrain but do not exclude a strong
breaking of the electroweak symmetry. The BESS model, on which we
have focused our discussion, has the advantage to provide a very
general frame leading in its minimal version to the existence of a
triplet of vector resonances to be discovered at 	LHC collider for
charged ones and at future linear $e^+e^- $ linear colliders for the
neutral one.

\end{document}